\documentclass[11pt]{myclass}
\usepackage{euscript}

\usepackage{amssymb}
\usepackage{epsfig}
\usepackage{xspace}
\usepackage{color}
\usepackage{url}
\usepackage{wrapfig}

\newcommand{\eps}{\varepsilon}

\newcommand{\Eu}[1]{\ensuremath{\EuScript{#1}}}
\newcommand{\bb}[1]{\ensuremath{\mathbb{#1}}}

\newcommand{\E}{\textbf{\textsf{E}}}
\newcommand{\Var}{\textbf{\textsf{Var}}}
\renewcommand{\Pr}{\textbf{\textsf{Pr}}}

\title{Chernoff-Hoeffding Inequality and Applications}
\author{Jeff M. Phillips}

\begin{document}
\maketitle


When dealing with modern big data sets, a very common theme is reducing the set through a random process.  These generally work by making ``many simple estimates'' of the full data set, and then judging them as a whole.  Perhaps magically, these ``many simple estimates'' can provide a very accurate and small representation of the large data set.  The key tool in showing how many of these simple estimates are needed for a fixed accuracy trade-off is the \emph{Chernoff-Hoeffding} inequality~\cite{Che52,Hoe63}.  This document provides a simple form of this bound, and two examples of its use.

\section{Chernoff-Hoeffding Inequality}  

We consider two specific forms of the Chernoff-Hoeffding bound.  They are not the strongest form of the bound, but is for many applications asymptotically equivalent, and it also fairly straight-forward to use.

\begin{theorem}
\label{thm:CH1}
Consider a set of $r$ \emph{independent} random variables $\{X_1, \ldots, X_r\}$.  If we know $a_i \leq X_i \leq b_i$, then let $\Delta_i = b_i - a_i$.  
Let $M = \sum_{i=1}^r X_i$.  Then for any $\alpha \in (0,1/2)$
\[
\Pr[| M - \E[M] | > \alpha] \leq 2 \exp\left( \frac{-2 \alpha^2}{\sum_{i=1}^r \Delta_i^2} \right).  
\]
\end{theorem}

\begin{theorem}
\label{thm:CH2}
Consider a set of $r$ independent identically distributed (iid) random variables $\{X_1, \ldots, X_r\}$ such that $-\Delta \leq X_i \leq \Delta$ and $\E[X_i] = 0$ for each $i \in [r]$.  Let $M = \sum_{i=1}^r X_i$ (a sum of $X_i$s).  Then for any $\alpha \in (0,1/2)$
\[
\Pr[|M| > \alpha] \leq 2 \exp\left( \frac{-\alpha^2}{2 r \Delta^2} \right).  
\]
\end{theorem}

We follow by stating a bound that depends only on the variance, but has an unfortunately quite strong requirement on $\alpha$.  We mainly state this form to help sketch the proof of an application to come.  

\begin{theorem}
\label{thm:CH}
Consider a set of $r$ \emph{independent} random variables $\{X_1, \ldots, X_r\}$.  
Let $M = \sum_{i=1}^r X_i$.  Then for $\alpha \in (0, 2 \Var[M] / (\max_i |X_i - \E[X_i]|) )$
\[
\Pr[| M - \E[M] | > \alpha] \leq 2 \exp\left( \frac{-\alpha^2}{4 \sum_{i=1}^r \Var[X_i]} \right).  
\]
\end{theorem}

\subsection{The Union Bound}
The \emph{Robin} to Chernoff-Hoeffding's \emph{Batman} is the \emph{union bound}.  It shows how to apply this single bound to many problems at once.  It may appear crude, but can usually only be significantly improved if special structure is available in the class of problems.  

\begin{theorem}
\label{thm:union}
Consider $t$ possibly \emph{dependent} random events $X_1, \ldots, X_t$.  The probability that all events occur is at least
\[
1 - \sum_{i=1}^t (1-\Pr[X_i]).
\]
\end{theorem}
That is, all events are true if no event is not true.

\section{Johnson-Lindenstrauss Lemma}  

The first example use is the Johnson-Lindenstrauss Lemma~\cite{JL84}.  It describes, in the worst case, how well are distances preserved under random projections.  
A random projection $\phi : \bb{R}^d \to \bb{R}^k$ can be defined by the $k$ independent (not necessarily orthogonal) coordinates, each expressed separately  $\phi_i : \bb{R}^d \to \bb{R}^1$ for $i \in [k]$.  
Specifically, $\phi_i$ is associated with an independent random vector $u_i \in \bb{S}^{d-1}$, that is a random unit vector in $\bb{R}^d$.  Then $\phi_i(p) = \langle p, u_i \rangle$, the inner (aka dot) product between $p$ and the random vector $u_i$.

\begin{theorem}[\cite{JL84}]
Consider a point set $P \subset \bb{R}^d$ of size $n$.  Let $Q = \phi(P)$ be a random linear projection of $P$ to $\bb{R}^k$ where $k = (8/\eps)^2 \ln(n/\delta)$.  Then with probability at least $1-\delta$ for \emph{all} $p, p' \in P$, and with $\eps \in (0,1/2]$
\begin{equation}
\label{eq:JL}
(1-\eps) \left\|p - p'\right\| \leq \sqrt{\frac{d}{k}} \left\|\phi\left(p\right) - \phi\left(p'\right)\right\| \leq (1+\eps) \left\|p - p'\right\|.
\end{equation}
\end{theorem}

To prove this we first note that the squared version of $\|\phi(p) - \phi(p')\|$ can be decomposed as follows:
\[
\|\phi(p) - \phi(p')\|^2 = \sum_{i=1}^k \|\phi_i(p) - \phi_i(p')\|^2.  
\]
Then since $(1-\eps) > (1-\eps)^2$ and $(1+\eps) < (1+\eps)^2$ for $\eps \in (0,1/2]$, it is sufficient and simpler to prove 
\begin{equation}
\label{eq:sqJL}
(1-\eps)  \leq \frac{d}{k} \frac{\|\phi(p) - \phi(p')\|^2}{\|p - p'\|^2} \leq (1+\eps).
\end{equation}

Now we consider the random variable $M = (d/k) \|\phi(p) - \phi(p')\|^2 / \|p - p'\|^2$ as the sum over $k$ random events $X_i =(d/k) \|\phi_i(p) - \phi_i(p')\|^2 / \|p - p'\|^2$.  
Now two simple observations follow:
\begin{itemize}
\item $\E[X_i] = 1/k$.  To see this, for each $u_i$ (independent of other $u_{i'}$, $i\neq i'$) consider a random rotation of the standard orthogonal basis, restricted only so that one axis is aligned to $u_i$ (which itself was random).  Then, in expectation each axis of this rotated basis contains $1/d$ of the squared norm of any vector, in particular $(p-p')$.  So $\E[(\langle u_i, p - p' \rangle)^2] = (1/d) \|p - p'\|^2$.  Then $\E[X_i] = 1/k$ follows from the linearity of $\phi$.  
\item $\Var[X_i] \leq 1/k^2$.  Since $\|\phi_i(p) - \phi(p')\|^2 \geq 0$, if the variance were larger than $1/k^2$, then the average distance from $E[X_i] = 1/k$ would be larger that $1/k$, and then the expected value would need to be larger than $1/k$.  
\end{itemize}

Now plugging these terms into Theorem \ref{thm:CH} yields (for some parameter $\gamma$)
\[
\Pr[|M - \E[M]| > \alpha] \leq 2 \exp\left(\frac{-\alpha^2}{4 k(1/k^2)} \right) \leq \gamma,
\]
and hence solving for $k$
\[
k \geq  4 \frac{1}{\alpha^2} \ln \left( \frac{2}{\gamma}\right).
\]
Set the middle term in (\ref{eq:sqJL}) to $M$ and note $\E[M] = 1$.  Now by setting $\alpha = \eps$\footnote{This is not the most general proof since Theorem \ref{thm:CH} requires $\eps = \alpha \leq 2 (k/k^2)/(d/k) = 2/d$ which is typically much smaller than $1/2$.}, it follows (\ref{eq:sqJL}) is satisfied with probability $1-\gamma$ for any one pair $p,p' \in P$ when $k \geq (4/\eps^2) \ln (2/\gamma)$.  
Since there are ${n \choose 2} < n^2$ pairs in $P$, by the union bound, setting $\gamma = \delta/n^2$ reveals that for $k \geq (8/\eps^2) \ln(n/\delta)$ ensures that \emph{all pairs} $p,p' \in P$ satisfy (\ref{eq:JL}) with probability at least $1-\delta$.  
\qed

There are several other (often more general) proofs of this theorem~\cite{FM88,IM98,DG03,Ach03,KM05,IN07,Mat08b}.

\newpage
\section{Subset Samples for Density Approximation}

\begin{wrapfigure}{r}[10pt]{5cm}
\hspace{-.8cm}
\includegraphics[width=\linewidth]{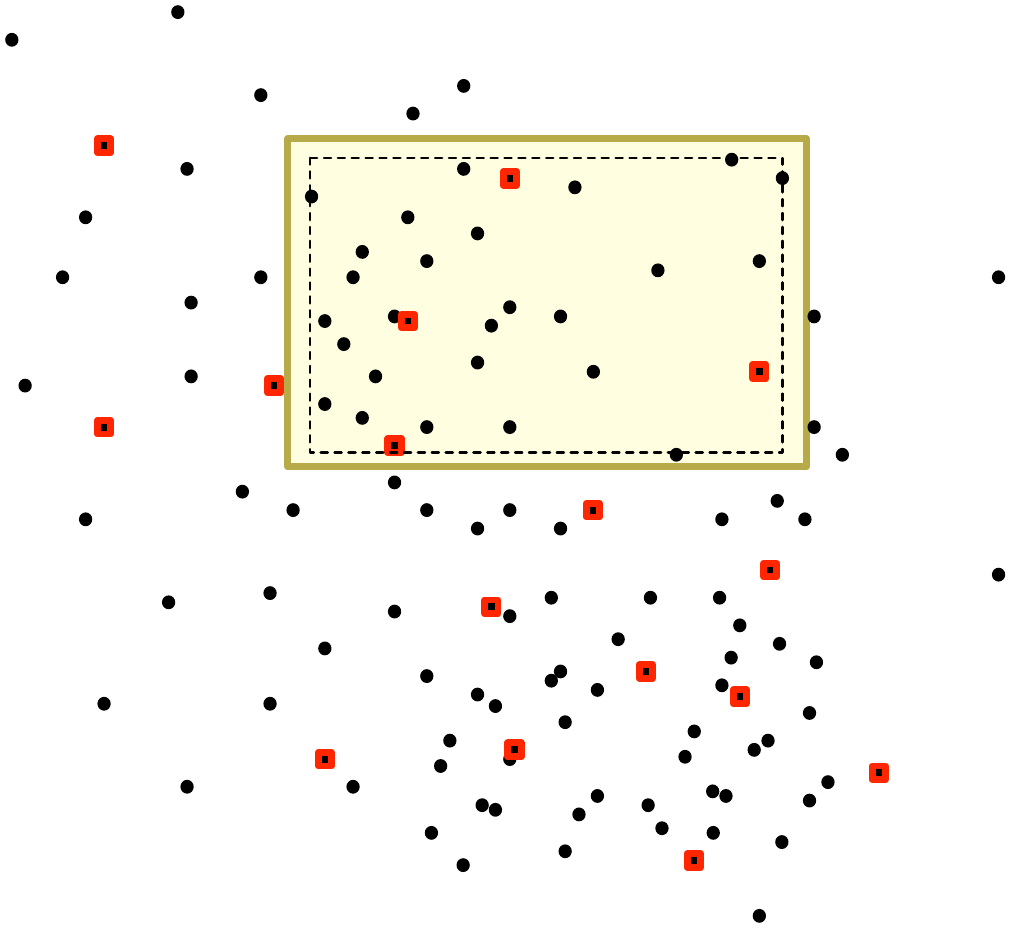}
\vspace{-.4cm}
\end{wrapfigure}
Again, consider a set of $n$ points $P \subset \bb{R}^d$.  Also consider a set $\Eu{R}$ of \emph{queries} we can ask on these points.  Herein let each $q \in \Eu{R}$ corresponds to a $d$-dimensional axis-aligned rectangle $R_q = [a_1, b_1] \times [a_2, b_2] \times \ldots \times [a_d, b_d]$, and asks for how many points in $P$ are in $R_q$.  That is $q(P) = |P \cap R_q|$.  
For example, if $P$ represents customers of a store with $d$ attributes (e.g. total number of purchases, average number of purchase each week, average purchase amount, \ldots)  and $R_q$ is a desired profile (e.g.  has between $100$ and $1000$ purchases total, averaging between $2.5$ and $10$ a week, with an average total purchase between $\$10$ and $\$20$, \ldots).  Then queries return the number of customers who fit that profile.  
This pair $(P,\Eu{R})$ is called a \emph{range space}.  

We now present a weak version of a theorem by Vapnik and Chervonenkis~\cite{VC71} about randomly sampling and range spaces.  

\begin{theorem}
\label{thm:eps-S}
Let $S \subset P$ be a random sample from $P$ of size $k = (d/\eps^2) \log (2n/\delta)$.  Then with probability at least $1-\delta$, for all $q \in \Eu{R}$
\begin{equation}
\label{eq:eps-S}
\left| \frac{q(P)}{|P|} - \frac{q(S)}{|S|} \right| \leq \eps.
\end{equation}
\end{theorem}

They key to this theorem is again the Chernoff-Hoeffding bound.  Fix some $q \in \Eu{R}$, and for each point $s_i$ in $S$, let $X_i$ be a random event describing the effect on $q(S)$ of $s_i$.  That is $X_i = 1$ if $s_i \in R_q$ and $X_i = 0$ if $s_i \notin R_q$, so $\Delta_i = 1$ for all $i \in [k]$.   
Let $M = \sum_i X_i = q(S)$, and note that $\E[M] = |S| \cdot q(P)/|P|$.  

Multiplying $M$ by $k = |S|$ we can now apply Theorem \ref{thm:CH1} to say
\[
\Pr\left[\left| \frac{q(S)}{|S|} - \frac{q(P)}{|P|} \right| \geq \eps \right]
= 
\Pr\left[\left| M - \E[M] \right| \geq \eps k \right] 
\leq 
2 \exp\left( \frac{- 2 (\eps k)^2}{\sum_{i=1}^k \Delta_i^2} \right)
=
2 \exp( -2 \eps^2 k )
\leq 
\gamma.
\]
Solving for $k$ yields that if $k \geq (1/2\eps^2) \ln (2/\gamma)$, then (\ref{eq:eps-S}) is true with probability at least $1-\gamma$ for our fixed $q \in \Eu{R}$.  

To extend this to all possible choices of $q \in \Eu{R}$ we need to apply the union bound on some bounded number of possible queries.  We can show that there are no more than $n^{2d}$ distinct subsets of $P$ that any axis-aligned-based query in $\Eu{R}$ can represent.  

To see this, take any rectangle $R$ that contains some subset of $T \subset P$ of the points in $P$.  Shrink this rectangle along each coordinate until no interval can be made smaller without changing the subset of points it contains.  At this point $R$ will touch at most $2d$ points, two for each dimension (if one side happens to touch two points simultaneously, this only lowers the number of possible subsets).  
Any rectangle can thus be mapped to one of at most $n^{2d}$ rectangles without changing which points it contains, where this \emph{canonical} rectangle (and importantly its subset of points) is described by this subset of $2d$ points.  

Since the application of the Chernoff-Hoeffding bound above does not change if the subset defined by $R_q$ does not change, to prove Theorem \ref{thm:eps-S} we need to show (\ref{eq:eps-S}) holds for only $n^{2d}$ different subsets.  Setting $\delta = \gamma/n^{2d}$ and apply the union bound (Theorem \ref{thm:union}) indicates that $k \geq (d/\eps^2) \ln (2n/\delta)$ random samples is sufficient.  \qed

\paragraph{Extensions:}
\begin{itemize} 
\item Amazingly, Vapnik and Chervonenkis~\cite{VC71} proved an ever stronger result that only $k = O((d/\eps^2) \log(1/\eps \delta))$ random samples are needed.  Note, this has no dependence on $n$, the number of points!  And moreover, Talagrand~\cite{Tal94}, as reported by Li, Long, and Srinivasan~\cite{LLS01} improved this further to $k = O((1/\eps^2) (d  + \log (1/\delta))$.  So basically the number of samples needed to guarantee any \emph{one} query has at most $\eps$-error, is sufficient to guarantee the same result for \emph{all} queries!

\item  This generalizes naturally to other types of range queries, using the idea of VC-dimension $\nu$; where the bound is then $k = O((1/\eps^2)(\nu + \log (1/\delta))$.  For axis-aligned rectangles $\nu = 2d$, for balls it is $\nu = d+1$, and for half spaces it is $\nu = d+1$.  This last bound for half spaces is particularly important for understanding how many samples are needed for determining approximate (linear) classifiers for machine learning.  

\item These bounds hold if $P$ is a continuous distribution (in some sense it has an infinite number of points).  
\end{itemize}

\section{Delayed Proofs}  

Here we prove Theorem \ref{thm:CH1}, inspired by the proof of Theorem 12.4 in Mitzenmacher and Upfal~\cite{MU05}.    
We then show Theorem \ref{thm:CH2} as a corollary.   
A proof of Theorem \ref{thm:CH} can be found in \cite{DP09}.

\paragraph{Markov inequality.}
Consider a random variable $X$ such that all possible values of $X$ are non-negative, then 
\[
\Pr[X > \alpha] \leq \frac{\E[X]}{\alpha}.
\]
To see this, consider if it was not true, and $\Pr[X > \alpha] > \E[X]/\alpha$.  Let $\gamma = \Pr[X > \alpha]$.  Then, since $X > 0$, we need to make sure the expected value of $X$ does not get too large.  So, let the instances of $X$ from the probability distribution of its values which are less than $\E[X]/\alpha$ be as small as possible, namely $0$.  Then we can still reach a contradiction:  
\[
\E[X] \geq (1 - \gamma) 0 + (\gamma) \alpha = \gamma \alpha > \frac{\E[X]}{\alpha} \alpha = \E[X].  
\]

\paragraph{Exponential inequalities.}
We state a simple fact about natural exponentials $e^x = \exp(x)$ that follows from its Taylor expansion.  
\begin{equation}
\label{eq:exp}
\frac{1}{2}(e^x + e^{-x}) \leq e^{x^2/2}
\end{equation}

\paragraph{Proof.}
We will prove the one-sided condition below.   The other side is symmetric, and the two-sided version follows from the union bound.  
\begin{equation}
\Pr[M - \E[M] > \alpha] \leq \exp \left(\frac{-2 \alpha^2}{\sum_{i=1}^r \Delta_i^2} \right).
\label{eq:one-side}
\end{equation}

We start by letting $Y_i = X_i - \E[X_i]$ and rewriting 
\begin{align*}
Y_i 
&= 
\Delta_i \frac{1 + Y_i/\Delta_i}{2} - \Delta_i \frac{1-Y_i/\Delta_i}{2}
\\
&= (\Delta_i) t + (-\Delta_i) (1-t),
\end{align*}
where $t = (1/2)(1+Y_i/\Delta_i)$; note that since $|Y_i| \leq \Delta_i$ then $t \in [0,1]$.  
Now since $e^{\lambda x}$ is convex in $x$ (we will set $\lambda = \alpha/\sum_{i=1}^r \Delta_i^2$ later), it follows that 
\begin{align*}
e^{\lambda Y_i}
& \leq 
e^{\lambda \Delta_i} \frac{1+Y_i/\Delta_i}{2} + e^{-\lambda \Delta_i} \frac{1 - Y_i/\Delta_i}{2}
\\ & = 
\frac{e^{\lambda \Delta_i} + e^{-\lambda \Delta_i}}{2} + \frac{Y_i}{2 \Delta_i} (e^{\lambda \Delta_i} + e^{-\lambda \Delta_i}).
\end{align*}
Now since $\E[Y_i] = 0$ and equation (\ref{eq:exp}) we have
\begin{equation}
\label{eq:toDelta}
\E\left[e^{\lambda Y_i}\right] 
\leq 
\E\left[\frac{e^{\lambda \Delta_i} + e^{-\lambda \Delta_i}}{2} + \frac{Y_i}{2 \Delta_i} (e^{\lambda \Delta_i} + e^{-\lambda \Delta_i}) \right]
= 
\frac{e^{\lambda \Delta_i} + e^{-\lambda \Delta_i}}{2}
\leq
\exp\left(\frac{\lambda^2 \Delta_i^2}{2}\right).
\end{equation}

Finally we can show equation (\ref{eq:one-side}) as follows
\begin{align*}
\Pr[M - \E[M] > \alpha] 
&=
\Pr \left[\sum_i (X_i - \E[X_i]) \geq \alpha \right] = \Pr \left[\sum_i Y_i \geq \alpha \right]
\\ &=
\Pr\left[\exp\left(\lambda \sum_i Y_i\right) > \exp(\lambda \alpha) \right]
\\ &\leq
\frac{1}{\exp(\lambda \alpha)} \E \left[\exp \left(\lambda \sum_i Y_i\right) \right]
=
\frac{1}{\exp(\lambda \alpha)}  \E \left[\prod_i \exp (\lambda Y_i) \right] 
\\ &\leq
\frac{1}{\exp(\lambda \alpha)}  \left(\prod_i \exp (\lambda^2 \Delta_i^2 / 2)\right) 
=
\exp \left(\frac{\lambda^2}{2} \sum_i \Delta_i^2 - \lambda \alpha \right)
\\ &=
\exp \left(\frac{-\alpha^2}{2 \sum_i \Delta_i^2} \right).
\end{align*}
The first inequality is from Markov inequality, the second from the equation (\ref{eq:toDelta}), and the last equality uses our choice of $\lambda = \alpha / \sum_i \Delta_i^2$.  
\qed

To see Theorem \ref{thm:CH2} from Theorem \ref{thm:CH1}, set each $\Delta_i = 2\Delta$ and $\E[M] = 0$.  

\subsection{On Independence and the Union Bound}
The proof of the union bound is an elementary observation.  Here we state a perhaps amazing fact that this seemingly crude bound is fairly tight even if the events are independent.  
Let $\Pr[X_i] = 1-\gamma$ for $i \in [t]$.  The union bound says the probability all events occur is at least $1-t \gamma$.  So to achieve a total of at most $\delta$ probability of failure, we need $\gamma \leq \delta/t$.  

On the other hand, by independence, we can state the probability of all events is $(1-\gamma)^t$.  By the approximation for large $s$ that $(1-x/s)^s \approx e^{-x}$ we can approximate
$(1-\gamma)^t \approx e^{-\gamma t}$.
So to achieve a total of at most $\delta$ probability of failure, we need  $1 - \delta \geq  e^{-\gamma t}$, which after some algebraic manipulation reveals 
$\gamma \leq \ln(1/(1-\delta)) / t$.   

So for $\delta$ small enough (say $\delta = 1/100$, then $\ln(1/(1-\delta)) = 0.01005\ldots$) the terms $\delta/t$ and $\ln(1/(1-\delta))/t$ are virtually the same.  
The only way to dramatically improve this is to show that the events are \emph{strongly negatively dependent}, as for instance is done in the proofs by Vapnik and Chervonenkis~\cite{VC71} and Talagrand~\cite{Tal94}.

\section*{Acknowledgements}
Thanks to Tomas Juskevicius and Graham Cormode for pointed out the extra conditions on $\alpha$ stated in what is now Theorem \ref{thm:CH}.  

\bibliographystyle{plain}
\bibliography{drefs}

\end{document}